\documentclass[12pt]{article}
\usepackage{amssymb,amsmath,epsfig}

\begin{document}

\title{\bf Energy Contents of Gravitational Waves in Teleparallel
Gravity}
\author{M. Sharif \thanks {msharif@math.pu.edu.pk} and Sumaira Taj
\thanks {sumairataj@ymail.com}\\
Department of Mathematics, University of the Punjab,\\
Quaid-e-Azam Campus, Lahore-54590, Pakistan.}

\date{}

\maketitle
\begin{abstract}
The conserved quantities, that are, gravitational energy-momentum
and its relevant quantities are investigated for cylindrical and
spherical gravitational waves in the framework of teleparallel
equivalent of General Relativity using the Hamiltonian approach. For
both cylindrical and spherical gravitational waves, we obtain
definite energy and constant momentum. The constant momentum shows
consistency with the results available in General Relativity and
teleparallel gravity. The angular momentum for cylindrical and
spherical gravitational waves also turn out to be constant. Further,
we evaluate their gravitational energy-momentum fluxes and
gravitational pressure.
\end{abstract}


\section{Introduction}

Gravitational waves are extremely elusive because gravity is a very
weak force.  These waves affect the matter, from which they pass
through, in a quite negligible manner which makes them difficult to
detect. Gravitational waves, by definition, have zero
energy-momentum tensor and hence their existence was questioned.
However, the theory of General Relativity (GR) predicts the
existence of gravitational waves, rising from the relation between
space and time. Indeed this problem arises because energy is not
well-defined in GR (as the strong equivalence principle refutes the
energy localization of the gravitational field). According to Synge
\cite{1}, energy of the gravitational field should be localizable
independent of any observer. Bondi \cite{2} argued that
non-localizable form of energy is inadmissible in GR. This
disputable point is the origin of a long-lasting discussion on the
energy and momentum transported by a gravitational wave.

Scheidegger \cite{3} raised question about the well-defined
existence of gravitational radiations. Ehlers and Kundt \cite{4}
resolved this problem for gravitational waves by analyzing a sphere
of test particles in the path of plane-fronted gravitational waves.
They showed that these particles acquired a constant momentum from
the waves. Weber and Wheeler \cite{5} gave the similar discussion
for cylindrical gravitational waves. Qadir and Sharif \cite{6}
explored an operational approach, embodying the same principle, to
show that gravitational waves impart momentum. One of us \cite{11}
found energy-momentum using various prescriptions that provide
acceptable results for different cosmological models and
gravitational waves.

It was suggested \cite{12,13} that the energy-momentum problem might
provide more better results in the framework of teleparallel
equivalent of General Relativity (TEGR). M{\o}ller \cite{14} was the
first who observed that the tetrad description of the gravitational
field could lead to a better expression for the gravitational
energy-momentum than does GR. Sharif and Nazir \cite{15}
investigated energy of cylindrical gravitational waves in GR and
teleparallel gravity and found inconsistence results for the two
theories. Andrade et al. \cite{16} considered the localization of
energy in Lagrangian framework of TEGR. Maluf et al. \cite{17}
derived an expression for the gravitational energy, momentum and
angular momentum using the Hamiltonian formulation of TEGR
\cite{18}. The same authors \cite{19} evaluated the energy flux of
gravitational waves in the framework of TEGR. Maluf and Ulhoa
\cite{20} showed that gravitational energy-momentum of plane-fronted
gravitational waves is non-positive.

In this paper, we use the Hamiltonian approach in TEGR to evaluate
energy and its contents for cylindrical and spherical gravitational
waves. In the next section, some basic concepts of TEGR and
energy-momentum expressions are given. Section \textbf{3} is devoted
for the evaluation of energy and its related quantities for
cylindrical gravitational waves. In section \textbf{4}, we calculate
these quantities for spherical gravitational waves. Summary and
discussion is presented in the last section.

\section{Hamiltonian Approach: Energy-Momentum in
Teleparallel Theory}

The Riemannian metric in terms of a non-trivial tetrad ${e^a}_\mu$
is written as
\begin{equation} \label{1}
g_{\mu\nu}=\eta_{ab} {e^a}_\mu {e^b}_\nu,
\end{equation}
where the tetrad field and its inverse satisfy the relation
\begin{equation}\label{2}
{e^a}_\mu{e_b}^\mu={\delta^a}_b,\quad
{e^a}_\mu{e_a}^\nu={\delta_\mu}^\nu.
\end{equation}
We denote the spacetime indices by Greek alphabets
$(\mu,\nu,\rho,...)$ and tangent space indices by Latin alphabets
$(a,b,c,...)$ and these run from 0 to 3. Time and space indices are
denoted according to $\mu=0,i,~a=(0),(i)$. The torsion tensor is
defined as
\begin{equation}\label{3}
{T^a}_{\mu\nu}=\partial_\mu{e^a}_\nu-\partial_\nu{e^a}_\mu,
\end{equation}
which is related to the Weitzenb\"{o}ck connection \cite{21}
\begin{equation}\label{4}
{\Gamma^\lambda}_{\mu\nu}={e_a}^\lambda\partial_\nu{e^a}_\mu.
\end{equation}

The Lagrangian density in TEGR is given by \cite{18}
\begin{eqnarray}\label{5}
L=-\kappa
e(\frac{1}{4}T^{abc}T_{abc}+\frac{1}{2}T^{abc}T_{bac}-T^{a}T_{a})-L_M
\equiv-\kappa e\Sigma^{abc}T_{abc}-L_M,
\end{eqnarray}
where $\kappa=1/16\pi$ and $e=det({e^a}_\mu)$. The tensor
$\Sigma^{abc}$ is defined as
\begin{equation}\label{6}
\Sigma^{abc}=\frac{1}{4}(T^{abc}+T^{bac}-T^{cab})
+\frac{1}{2}(\eta^{ac}T^b-\eta^{ab}T^c)
\end{equation}
which satisfies the antisymmetric property, i.e.,
$\Sigma^{abc}=-\Sigma^{acb}.~L_M$ denotes the Lagrangian density for
the matter fields, $T_a={T^b}_{ba}$ and $T_{abc}={e_b}^\mu{e_c}^\nu
T_{a\mu\nu}$. The variation of the Lagrangian $L$ with respect to
$e^{a\mu}$ yields the field equations
\begin{equation}\label{7}
e_{a\lambda}e_{b\mu}\partial_\nu(e\Sigma^{b\lambda\nu})
-e({\Sigma^{b\nu}}_a T_{b\nu\mu}-\frac{1}{4}e_{a\mu}
T_{bcd}\Sigma^{bcd})=\frac{1}{4\kappa}e T_{a\mu},
\end{equation}
where
\begin{equation*}
\frac{\delta L_M}{\delta e^{a\mu}}=e T_{a\mu}.
\end{equation*}
The total Hamiltonian density is \cite{22}
\begin{equation}\label{8}
H(e_{ai},\Pi_{ai})=e_{a0}C^a+\alpha_{ik}\Gamma^{ik}+\beta_k\Gamma^k
+\partial_k(e_{a0}\Pi^{ak}),
\end{equation}
where $C^a,~\Gamma^{ik}$ and $\Gamma^k$ are primary constraints,
$\alpha_{ik}$ and $\beta_k$ are the Lagrangian multipliers defined
as $\alpha_{ik}=\frac{1}{2}(T_{i0k}+T_{k0i})$ and $\beta_k=T_{00k}$.
$C^a$ is given by a total divergence in the form
$C^a=-\partial_i\Pi^{ai}+H^a$, where
\begin{equation}\label{10}
\Pi^{ai}=-4\kappa e\Sigma^{a0i}
\end{equation}
is the momentum canonically conjugated to $e_{ai}$. The term
$-\partial_i\Pi^{ai}$ is identified as the \textit{energy-momentum
density} \cite{17}. The total \textit{energy-momentum} is defined
as
\begin{equation}\label{11}
P^a=-\int_V d^3 x\partial_i\Pi^{ai},
\end{equation}
where $V$ is an arbitrary space volume.

The constraint
\begin{eqnarray*}
\Gamma^{ik}=-\Gamma^{ki}=2\Pi^{[ik]}-2\kappa
e[-g^{im}g^{kj}{T^0}_{mj}+(g^{im}g^{0k}-g^{km}g^{0i}){T^j}_{mj}]=0
\end{eqnarray*}
gives
\begin{equation}\label{12}
2\Pi^{[ik]}=2\kappa
e[-g^{im}g^{kj}{T^0}_{mj}+(g^{im}g^{0k}-g^{km}g^{0i}){T^j}_{mj}]
\end{equation}
which is referred to as the \textit{angular momentum density}.
Consequently, the \textit{angular momentum} is defined as
\begin{eqnarray}\label{13}
M^{ik}=2\int_V d^3 x\Pi^{[ik]}=2\kappa\int_V d^3x
e[-g^{im}g^{kj}{T^0}_{mj}+(g^{im}g^{0k}-g^{km}g^{0i}){T^j}_{mj}].
\end{eqnarray}
We can write using the field equations as
\begin{equation}\label{16}
\frac{d}{dt}[-\int_V d^3x\partial_j\Pi^{aj}]=-\Phi^a_g-\Phi^a_m,
\end{equation}
where
\begin{equation}\label{17}
\Phi^a_g=\int_S dS_j\phi^{aj},\quad \Phi^a_m=\int_S dS_j(e{e^a}_\mu
T^{j\mu})
\end{equation}
are the $a$ components of the \textit{gravitational and matter
energy-momentum flux}. The quantity
\begin{equation}\label{19}
\phi^{aj}=\kappa e e^{a\mu}(4\Sigma^{bcj}T_{bc\mu}
-\delta^j_\mu\Sigma^{bcd}T_{bcd})
\end{equation}
represent the $a$ component of the gravitational energy-momentum
flux density in $j$ direction. In terms of the gravitational
energy-momentum, Eq.(\ref{16}) takes the form
\begin{equation}\label{20}
\frac{dP^a}{dt}=-\Phi^a_g-\Phi^a_m.
\end{equation}

For the vacuum spacetime, the above equation reduces to
\begin{equation}\label{21}
\frac{dP^a}{dt}=-\Phi^a_g=-\int_S dS_j\phi^{aj}.
\end{equation}
If we take $a=(i)=(1),(2),(3)$, then
\begin{equation}\label{22}
\frac{dP^{(i)}}{dt}=\int_S dS_j(-\phi^{(i)j}).
\end{equation}
The left hand side of the above equation has the character of force
while the density $(-\phi^{(i)j})$ is considered as a force per unit
area, or pressure density. Thus Eq.(\ref{22}) has the nature of the
gravitational pressure.

\section{Cylindrical Gravitational Waves}

The line element of cylindrical gravitational waves given by
Einstein and Rosen is \cite{5}
\begin{equation}\label{27}
ds^2=-e^{2(\gamma-\psi)}dt^2+e^{2(\gamma-\psi)}d\rho^2
+\rho^2e^{-2\psi}d\phi^2+e^{2\psi}dz^2,
\end{equation}
where the arbitrary functions,
$\gamma=\gamma(\rho,t),~\psi=\psi(\rho,t)$, satisfy the vacuum field
equations
\begin{eqnarray}\label{28}
\psi''+\frac{1}{\rho}\psi'-\ddot{\psi}=0,\quad
\gamma'=\rho(\psi'^2+\dot{\psi}^2),\quad
\dot{\gamma}=2\rho\dot{\psi}\psi',
\end{eqnarray}
dot and prime represent differentiation with respect to $t$ and
$\rho$ respectively. The tetrad field, satisfying Eqs.(\ref{1}) and
(\ref{2}) is
\begin{equation}\label{29}
{e^a}_\mu(t,\rho,\phi)=\left(
\begin{array}{cccc}
A & 0 & 0 & 0 \\
0 & A\cos\phi & -\rho C\sin\phi & 0 \\
0 & A\sin\phi & \rho C\cos\phi & 0 \\
0 & 0 & 0 & B \\
\end{array}
\right).
\end{equation}
Here $A=e^{(\gamma-\psi)},~B=e^\psi,~C=e^{-\psi}$ and its
determinant is
\begin{equation*}
e=det({e^a}_\mu)=\rho A^2.
\end{equation*}
The non-zero components of the torsion tensor are
\begin{eqnarray}\label{30}
T_{(0)01}&=&A',\quad T_{(1)01}=\dot{A}\cos\phi,\quad
T_{(1)02}=-\rho\dot{C}\sin\phi,\nonumber\\
T_{(1)12}&=&(A-C-\rho C')\sin\phi,\quad
T_{(2)01}=\dot{A}\sin\phi,\quad
T_{(2)02}=\rho\dot{C}\cos\phi,\nonumber\\
T_{(2)12}&=&-(A-C-\rho C')\cos\phi,\quad T_{(3)03}=\dot{B},\quad
T_{(3)13}=B'
\end{eqnarray}
which give rise to
\begin{eqnarray}\label{31}
T_{001}&=&AA',\quad T_{101}=A\dot{A},\qquad T_{202}
=\rho^2C\dot{C},\nonumber\\
T_{212}&=&\rho C(C-A+\rho C'),\quad T_{303}=B\dot{B}, \quad
T_{313}=BB'.
\end{eqnarray}

\subsection{Energy, Momentum and Angular Momentum}

The components of energy-momentum density, $-\partial_i\Pi^{ai}$,
for the cylindrical gravitational waves are found by using
Eqs.(\ref{6}) and (\ref{10})
\begin{eqnarray}\label{32}
-\partial_i\Pi^{(0)i}&=&-\partial_1[-2\kappa e^{(\psi-\gamma)}
(e^\gamma-1)]=2\kappa\partial_1[e^{(\psi-\gamma)}
(e^\gamma-1)],\nonumber\\
-\partial_i\Pi^{(1)i}&=&-\partial_2(2\kappa
\dot{\gamma}e^\psi\sin\phi)=-2\kappa\partial_2(\dot{\gamma}
e^\psi\sin\phi),\nonumber\\
-\partial_i\Pi^{(2)i}&=&-\partial_2(-2\kappa\dot{\gamma}e^\psi
\cos\phi)=2\kappa\partial_2(\dot{\gamma}e^\psi\cos\phi),\nonumber\\
-\partial_i\Pi^{(3)i}&=&-\partial_3[-2\kappa\rho e^{-\psi}
(\dot{\gamma}-2\dot{\psi})]=2\kappa\partial_3[\rho e^{-\psi}
(\dot{\gamma}-2\dot{\psi})].
\end{eqnarray}
Using these values in Eq.(\ref{11}) and integration over a
cylindrical region of an arbitrary length $L$ and radius $\rho$
gives energy and momentum
\begin{equation}\label{33}
P^{(0)}=\frac{1}{4}L e^{(\psi-\gamma)}(e^\gamma-1),\quad P^{(i)}=0.
\end{equation}
The angular momentum for cylindrical gravitational waves becomes
constant.

\subsection{Energy-Momentum Flux}

The components of gravitational energy flux density for cylindrical
gravitational waves, obtained by using Eq.(\ref{19}), are
\begin{equation}\label{35}
\phi^{(0)1}=-2\kappa\dot{\psi}e^{(\psi-\gamma)}(2\psi'\rho+e^\gamma-1),
\quad\phi^{(0)2}=0=\phi^{(0)3}.
\end{equation}
Consequently, the gravitational energy flux becomes
\begin{equation}\label{36}
\Phi^{(0)}_g=-\frac{1}{4}L e^{(\psi-\gamma)}
\{\dot{\psi}(e^\gamma-1)+\dot{\gamma}\}+\textmd{constant}.
\end{equation}
The components of gravitational momentum flux
\begin{eqnarray}\label{37}
\Phi^{(1)}_g&=&-2\kappa L\sin\phi\int e^\psi(\dot{\psi}^2-\psi'^2)
(1-\rho\psi')d\rho+\textmd{constant},\nonumber\\
\Phi^{(2)}_g&=&2\kappa L\cos\phi\int e^\psi(\dot{\psi}^2-\psi'^2)
(1-\rho\psi')d\rho+\textmd{constant},\nonumber\\
\Phi^{(2)}_g&=&\frac{1}{4}\int e^{-\psi}\{\rho^2\psi'(\dot{\psi}^2-\psi'^2)
+2\rho\psi'^2+\psi'(e^\gamma-1)\}d\rho\nonumber\\
&&-\frac{1}{4}e^{(\gamma-\psi)}+\textmd{constant}
\end{eqnarray}
are obtained by making use of the components of gravitational
momentum flux densities
\begin{equation}\label{38}
\phi^{(1)1}=2\kappa\gamma'e^\psi\cos\phi,\quad
\phi^{(1)2}=-2\kappa e^\psi\sin\phi(\dot{\psi}^2-\psi'^2)
(1-\rho\psi'),\quad\phi^{(1)3}=0,
\end{equation}
\begin{equation}\label{39}
\phi^{(2)1}=2\kappa\gamma'e^\psi\sin\phi,\quad \phi^{(2)2}=2\kappa
e^\psi\cos\phi(\dot{\psi}^2-\psi'^2)
(1-\rho\psi'),\quad\phi^{(2)3}=0,
\end{equation}
\begin{equation}\label{40}
\phi^{(3)1}=0=\phi^{(3)2},\quad
\phi^{(3)3}=2\kappa e^{-\psi}\{\rho^2\psi'(\dot{\psi}^2-\psi'^2)
+2\rho\psi'^2+\psi'(e^\gamma-1)-\gamma'e^\gamma\}
\end{equation}
in Eq.(\ref{19}) for $a=(1),(2)$ and $(3)$ respectively. Here the
energy-momentum flux represents the transfer of energy-momentum in
cylindrical gravitational waves. Notice that Maluf et al. \cite{19}
also found exactly the same energy but slightly different energy
flux.

\subsection{Gravitational Pressure}

In order to calculate gravitational pressure for cylindrical
gravitational waves, we use Eq.(\ref{22}) and confine the
considerations to a surface along radial direction, i.e.,
\begin{equation}\label{41}
\frac{dP^{(i)}}{dt}=\int_S dS_1(-\phi^{(i)1}).
\end{equation}
Using the components of gravitational momentum
flux density $\phi^{(i)1}$
\begin{equation}\label{42}
\phi^{(1)1}=2\kappa\cos\phi(\gamma'e^\psi),\quad
\phi^{(2)1}=2\kappa\sin\phi(\gamma'e^\psi),\quad\phi^{(3)1}=0,
\end{equation}
in Eq.(\ref{41}) and taking the unit
vector $\hat{\textbf{r}} =(\cos\phi,\sin\phi,0)$, it follows that
\begin{equation}\label{43}
\frac{d\textbf{P}}{dt}=-2\kappa(\gamma'e^\psi) \int_S d\phi
dz\hat{\textbf{r}}.
\end{equation}
Conversion of surface element $d\phi dz$ into spherical polar
coordinates, we have
\begin{equation}\label{44}
\frac{d\textbf{P}}{dt}=-2\kappa(\gamma'e^\psi)
\int_S\rho\sin\theta d\theta d\phi\hat{\textbf{r}}.
\end{equation}
Integration over a small solid angle $d\Omega=\sin\theta d\theta
d\phi$ of constant radius $\rho$ gives
\begin{equation}\label{45}
\frac{d\textbf{P}}{dt}=-2\kappa(\rho\gamma'e^\psi)
\Delta\Omega\hat{\textbf{r}}.
\end{equation}
Replacing $dt\rightarrow
d(ct),~\kappa=\frac{1}{16\pi}\rightarrow\frac{c^3}{16\pi G}$ in
the above equation, we obtain
\begin{equation}\label{46}
\frac{d\textbf{P}}{dt}=-(\gamma'e^\psi)\frac{c^4}{8\pi G\rho}
(\rho^2\Delta\Omega)\hat{\textbf{r}}.
\end{equation}
The quantity $-(\gamma'e^\psi){c^4}/{8\pi G\rho}$ on the right hand
side of this equation gives the gravitational pressure exerted on
the area element $(\rho^2\Delta\Omega)$. This equation can also be
written as
\begin{equation}\label{47}
\frac{d}{dt}\left(\frac{\textbf{P}}{M}\right)=-(\gamma'e^\psi)
\frac{c^4}{8\pi GM}\rho\Delta\Omega\hat{\textbf{r}}.
\end{equation}
The left hand side corresponds to acceleration which can be taken as
the gravitational acceleration field acting on the solid angle
$\Delta\Omega$ at a radial distance $\rho$.

\section{Spherical Gravitational Waves}

The line element describing the gravitational waves with spherical
wavefronts is \cite{25}
\begin{equation}\label{48}
ds^2=e^{-M}(-dt^2+d\rho^2)+e^{-U}(e^{-V}d\phi^2+e^Vdz^2),
\end{equation}
where $M,~U$ and $V$ are arbitrary functions depending on $t$ and
$\rho$. The vacuum field equations imply that $e^{-U}$ and $V$
satisfy the wave equation
\begin{equation}\label{49}
(e^{-U})_{tt}-(e^{-U})_{\rho\rho}=0,\quad
V_{tt}-U_tV_t-V_{\rho\rho}+U_\rho V_\rho=0.
\end{equation}
Equations for $M$ are
\begin{eqnarray}\label{51}
U_{tt}-U_{\rho\rho}&=&\frac{1}{2}(U_t^2+U_\rho^2+V_t^2+V_\rho^2)
-U_tM_t-U_\rho M_\rho=0,\\
2U_{t\rho}&=&U_tU_\rho-U_tM_\rho-U_\rho M_t+V_tV_\rho.\label{52}
\end{eqnarray}
The tetrad field corresponding to the metric (\ref{48}) is
\begin{equation}\label{53}
{e^a}_\mu(t, \rho,\phi)=\left(\begin{array}{cccc}
e^{-\frac{M}{2}} & 0 & 0 & 0 \\
0 & e^{-\frac{M}{2}}\cos\phi & -e^{\frac{(-U-V)}{2}}\sin\phi & 0 \\
0 & e^{-\frac{M}{2}}\sin\phi & e^{\frac{(-U-V)}{2}}\cos\phi & 0 \\
0 & 0 & 0 & e^{\frac{(-U+V)}{2}} \\
\end{array}
\right)
\end{equation}
and its determinant is $e=e^{(-M-U)}$. The non-vanishing components
of the torsion tensor are
\begin{eqnarray}\label{54}
T_{(0)01}&=&-\frac{M'}{2}e^{-\frac{M}{2}},\quad T_{(1)01}=-\frac{\dot{M}}{2}\cos\phi,\quad
T_{(1)02}=\frac{1}{2}(\dot{U}+\dot{V})e^{\frac{(-U-V)}{2}}\sin\phi,\nonumber\\
T_{(1)12}&=&\sin\phi\{\frac{1}{2}(U'+V')e^{\frac{(-U-V)}{2}}+e^{-\frac{M}{2}}\},\quad
T_{(2)01}=-\frac{\dot{M}}{2}e^{-\frac{M}{2}}\sin\phi,\nonumber\\
T_{(2)02}&=&-\frac{1}{2}(\dot{U}+\dot{V})e^{\frac{(-U-V)}{2}}\cos\phi,\quad
T_{(3)03}=\frac{1}{2}(\dot{V}-\dot{U})e^{\frac{(-U+V)}{2}},\nonumber\\
T_{(2)12}&=&-\cos\phi\{\frac{1}{2}(U'+V')e^{\frac{(-U-V)}{2}}+e^{-\frac{M}{2}}\},\quad
T_{(3)13}=\frac{1}{2}(V'-U')e^{\frac{(-U+V)}{2}}.\nonumber\\
\end{eqnarray}
The tensor $T_{\lambda\mu\nu}= {e^a}_\lambda T_{a\mu\nu}$ becomes
\begin{eqnarray}\label{55}
T_{001}&=&-\frac{M'}{2}e^{-M},\quad T_{101}=-\frac{\dot{M}}{2}e^{-M},
\quad T_{202}=-\frac{1}{2}(\dot{U}+\dot{V})e^{\frac{(-U-V)}{2}},\nonumber\\
T_{212}&=&-\{\frac{1}{2}(U'+V')e^{(-U-V)}+e^{-\frac{M}{2}}e^{\frac{(-U-V)}{2}}\},\nonumber\\
T_{303}&=&\frac{1}{2}(\dot{V}-\dot{U})e^{(-U+V)},\quad
T_{313}=\frac{1}{2}(V'-U')e^{(-U+V)}.
\end{eqnarray}

\subsection{Energy, Momentum and Angular Momentum}

The components of energy-momentum density for the spherical
gravitational waves become
\begin{eqnarray}\label{56}
-\partial_i\Pi^{(0)i}&=&2\kappa{\partial_1}\{e^{-U}
(e^{\frac{(U+V)}{2}}+U'e^{\frac{M}{2}})\},\nonumber\\
-\partial_i\Pi^{(1)i}&=&-\kappa e^{-U}\cos\phi\{2e^{\frac{M}{2}}
(\dot{U'}-\dot{U}U'+\frac{\dot{U}M'}{2})+e^{\frac{(U+V)}{2}}
(\dot{V}-\dot{U}-\dot{M})\},\nonumber\\
-\partial_i\Pi^{(2)i}&=&-\kappa e^{-U}\sin\phi\{2e^{\frac{M}{2}}
(\dot{U'}-\dot{U}U'+\frac{\dot{U}M'}{2})+e^{\frac{(U+V)}{2}}
(\dot{V}-\dot{U}-\dot{M})\},\nonumber\\
-\partial_i\Pi^{(3)i}&=&0.
\end{eqnarray}
Inserting these values in Eq.(\ref{11}), we obtain energy and
momentum as
\begin{equation}\label{57}
P^{(0)}=\frac{1}{4}Le^{-U}(e^{\frac{(U+V)}{2}}+U'e^{\frac{M}{2}}),\quad
P^{(i)}=0.
\end{equation}
Again, all the components of angular momentum turn out to be
constant.

\subsection{Energy-Momentum Flux}

The components of gravitational energy flux density
\begin{equation}\label{59}
\phi^{(0)1}=-\kappa\{e^{\frac{M}{2}}e^{-U}(V'\dot{V}-U'\dot{U}-M'\dot{U})
+e^{\frac{(-U+V)}{2}}(\dot{V}-\dot{U})\},\quad\phi^{(0)2}=0=\phi^{(0)3}
\end{equation}
give rise to gravitational energy flux
\begin{equation}\label{60}
\Phi^{(0)}_g=-\frac{1}{8}L\{e^{\frac{M}{2}}e^{-U}(V'\dot{V}-U'\dot{U}-M'\dot{U})
+e^{\frac{(-U+V)}{2}}(\dot{V}-\dot{U})\}+\textmd{constant}.
\end{equation}
Inserting the components of gravitational flux densities
\begin{eqnarray}\label{61}
\phi^{(1)1}&=&\kappa e^{\frac{M}{2}}e^{-U}\cos\phi\{(V'-U'-M')
e^{-\frac{M}{2}}e^{\frac{(U+V)}{2}}-\dot{U}^2-U'^2+\dot{U}\dot{M}\},\nonumber\\
\phi^{(1)2}&=&-\frac{1}{2}\kappa e^{\frac{(-U+V)}{2}}\sin\phi\{
\dot{M}(\dot{V}-\dot{U})+M'(U'-V')\},\quad \phi^{(1)3}=0,\\
\label{62}
\phi^{(2)1}&=&\kappa
e^{\frac{M}{2}}e^{-U}\sin\phi\{(V'-U'-M')
e^{-\frac{M}{2}}e^{\frac{(U+V)}{2}}-\dot{U}^2-U'^2+\dot{U}\dot{M}\},\nonumber\\
\phi^{(2)2}&=&\frac{1}{2}\kappa e^{\frac{(-U+V)}{2}}\cos\phi\{
\dot{M}(\dot{V}-\dot{U})+M'(U'-V')\},\quad\phi^{(2)3}=0,
\\
\label{63}
\phi^{(3)1}&=&0=\phi^{(3)2},\nonumber\\
\phi^{(3)3}&=&\frac{1}{2}\kappa[2M'e^{-\frac{M}{2}}+e^{\frac{(-U-V)}{2}}
\{M'(U'+V')-\dot{M}(\dot{V}+\dot{U})\}]
\end{eqnarray}
in Eq.(14), we obtain the gravitational momentum flux
\begin{eqnarray}\label{64}
\Phi^{(1)}_g&=&-\frac{1}{2}\kappa L\sin\phi\int e^{\frac{(-U+V)}{2}}
\{\dot{M}(\dot{V}-\dot{U})+M'(U'-V')\}d\rho+\textmd{constant},\nonumber\\
\Phi^{(2)}_g&=&\frac{1}{2}\kappa L\cos\phi\int e^{\frac{(-U+V)}{2}}
\{\dot{M}(\dot{V}-\dot{U})+M'(U'-V')\}d\rho+\textmd{constant},\nonumber\\
\Phi^{(3)}_g&=&-\frac{1}{4}e^{-\frac{M}{2}}+\textmd{constant}+\frac{1}{16}
\int e^{\frac{(-U-V)}{2}}\{M'(U'+V')-\dot{M}(\dot{U}+\dot{V})\}d\rho.\nonumber\\
\end{eqnarray}

\subsection{Gravitational Pressure}

For spherical gravitational waves, the gravitational momentum
flux density components become
\begin{eqnarray}\label{65}
\phi^{(1)1}&=&\kappa e^{\frac{M}{2}}e^{-U}\cos\phi\{(V'-U'-M')
e^{-\frac{M}{2}}e^{\frac{(U+V)}{2}}-\dot{U}^2-U'^2+\dot{U}\dot{M}\},\nonumber\\
\phi^{(2)1}&=&\kappa e^{\frac{M}{2}}e^{-U}\sin\phi\{(V'-U'-M')
e^{-\frac{M}{2}}e^{\frac{(U+V)}{2}}-\dot{U}^2-U'^2+\dot{U}\dot{M}\},\nonumber\\
\phi^{(3)1}&=&0.
\end{eqnarray}
Substituting these values in Eq.(\ref{41}) and taking the unit
vector $\hat{\textbf{r}} =(\cos\phi,\sin\phi,0)$, it follows that
\begin{equation}\label{66}
\frac{d\textbf{P}}{dt}=-\kappa e^{\frac{M}{2}}e^{-U}\{(V'-U'-M')
e^{-\frac{M}{2}}e^{\frac{(U+V)}{2}}-\dot{U}^2-U'^2+\dot{U}\dot{M}\}\int_S d\phi
dz\hat{\textbf{r}}.
\end{equation}
Proceeding in a similar way as for the cylindrical gravitational
waves, it follows that
\begin{eqnarray}\label{68}
\frac{d\textbf{P}}{dt}&=&\left[-\frac{c^4}{16\pi\rho
G}e^{\frac{M}{2}}e^{-U}\{(V'-U'-M')
e^{-\frac{M}{2}}e^{\frac{(U+V)}{2}}-\dot{U}^2-U'^2+\dot{U}\dot{M}\}\right]\nonumber\\
&\times&\hat{\textbf{r}}(\rho^2\Delta\Omega).
\end{eqnarray}
The term in the square brackets on the right hand side of the above
equation is interpreted as the gravitational pressure exerted on the
area element $(\rho^2\Delta\Omega)$. Equation (\ref{68}) can be
re-written as
\begin{equation}\label{69}
\frac{d}{dt}\left(\frac{\textbf{P}}{M}\right)
=\frac{c^4}{16\pi GM}e^{\frac{M}{2}}e^{-U}\{(V'-U'-M')
e^{-\frac{M}{2}}e^{\frac{(U+V)}{2}}
-\dot{U}^2-U'^2+\dot{U}\dot{M}\}\rho\Delta\Omega\hat{\textbf{r}}.
\end{equation}
The left hand side of this equation can be recognized as the
gravitational acceleration. We can consider it as the
gravitational acceleration field that acts on the solid angle
$\Delta\Omega$.

\section{Summary and Discussion}

In this paper, we have applied the coordinate independent
prescription obtained by using the Hamiltonian approach in TEGR to
investigate energy-momentum distribution of gravitational waves. We
have evaluated energy, momentum, angular momentum, gravitational
energy-momentum flux and gravitational pressure of cylindrical and
spherical gravitational waves. For cylindrical gravitational waves,
the energy expression turns out to be definite and well defined. The
constant momentum corresponds to the result of GR \cite{26} and
teleparallel gravity \cite{27}. In the case of spherical
gravitational waves, we obtain well-defined energy and constant
momentum which corresponds to the result of GR \cite{11}. The
angular momentum for these solutions turns out to be constant.

It is interesting to note that for cylindrical gravitational waves,
we obtain $P^aP^b\eta_{ab}=0$ if $\gamma=0$ which depicts the
property of a plane electromagnetic wave. If we take the background
region of spherical waves ($t<\rho$, Minkowski) described by the
solution $U=-\ln t-\ln\rho, V=\ln t-\ln\rho$ and $M=0$, Eq.(47)
yields energy-momentum zero while gravitational energy-momentum flux
becomes constant as expected. This is what one can expect for
Minkowski spacetime. Further, we have also evaluated the
gravitational pressure exerted by gravitational waves (cylindrical
and spherical). This may be helpful to investigate the
thermodynamics of the gravitational field.

We would like to mention here that our results show consistency with
the results of different energy-momentum complexes both in GR and
teleparallel gravity. Here we can express these conserved quantities
such as energy, momentum and angular momentum tensor of the
gravitational field covariantly. Finally, we can say that the tetrad
formulism provides a more satisfactory treatment of the localization
problem.

\section*{Appendix}
The non-zero components of the tensor
\begin{equation*}
\Sigma^{abc}=\frac{1}{4}(T^{abc}+T^{bac}-T^{cab})
+\frac{1}{2}(\eta^{ac}T^b-\eta^{ab}T^c),
\end{equation*}
for
\begin{itemize}
\item \textbf{Cylindrical Gravitational Waves}
\begin{eqnarray*}
\Sigma^{001}&=&\frac{1}{2\rho}e^{4(\psi-\gamma)}(e^\gamma-1),\quad
\Sigma^{202}=\frac{1}{2\rho^2}\dot{\gamma}e^{2(2\psi-\gamma)},\\
\Sigma^{212}&=&-\frac{1}{2\rho^2}\gamma'e^{2(2\psi-\gamma)},\quad
\Sigma^{303}=\frac{1}{2}e^{-2\gamma}(\dot{\gamma}-2\dot{\psi}),\\
\Sigma^{313}&=&\frac{1}{2\rho}e^{-2\gamma}\{\rho(2\psi'-\gamma')
+e^\gamma-1\},
\end{eqnarray*}
\item \textbf{Spherical Gravitational Waves}
\begin{eqnarray*}
\Sigma^{001}&=&\frac{1}{2}e^{2M}(e^{\frac{(U+V)}{2}}e^{-\frac{M}{2}}+U'),
\quad\Sigma^{101}=-\frac{1}{2}\dot{U}e^{2M},\\
\Sigma^{202}&=&\frac{1}{4}e^{M}e^{(U+V)}(\dot{V}-\dot{M}-\dot{U}),\\
\Sigma^{212}&=&\frac{1}{4}e^{M}e^{(U+V)}(U'-V'+M'),\\
\Sigma^{303}&=&-\frac{1}{4}e^{M}e^{(U-V)}(\dot{V}+\dot{M}+\dot{U}),\\
\Sigma^{313}&=&\frac{1}{4}e^{M}e^{(U-V)}(U'+V'+M'+2e^{\frac{(U+V)}{2}}e^{-\frac{M}{2}}).
\end{eqnarray*}
\end{itemize}

\end{document}